\documentclass[12pt,a4paper]{article}
\usepackage{axodraw}
\usepackage{graphicx}
\usepackage{amsmath}
\usepackage{amsfonts}
\usepackage{amssymb}
\newcommand{\vs}{\vspace{-0.25cm}}

\newcommand{\fmd}{\,\mathrm{fm}^{-3}}

\begin{document}
\hfill TUM/T39-02-??

\begin{center}

{\Large
\textbf{Single-particle potential in a chiral approach to \\ 
nuclear matter including short range NN-terms}\footnote{
Work supported in part by BMBF, GSI and DFG.} 
}

\bigskip

\bigskip
S. Fritsch and N. Kaiser\\

\bigskip

{\small Physik Department T39, Technische Universit\"{a}t M\"{u}nchen, D-85747
Garching, Germany\\

\smallskip

{\it email: nkaiser@physik.tu-muenchen.de}}

\end{center}

\bigskip

\begin{abstract}
We extend a recent chiral approach to nuclear matter of Lutz et al. 
[Phys. Lett. {\bf B474} (2000) 7] by calculating the underlying
(complex-valued) single-particle potential $U(p,k_f)+i\,W(p,k_f)$. The 
potential for a nucleon at the bottom of the Fermi-sea, $U(0,k_{f0})= - 20.0
$\,MeV, comes out as much too weakly attractive in this approach. Even more 
seriously, the total single-particle energy does not rise monotonically with
the nucleon momentum $p$, implying a negative effective nucleon mass at the
Fermi-surface. Also, the imaginary single-particle potential, $W(0,k_{f0})=
51.1$\,MeV, is too large. More realistic single-particle properties together 
with a good nuclear matter equation of state can be obtained if the short range
contributions of non-pionic origin are treated in mean-field approximation
(i.e. if they are not further iterated with $1\pi$-exchange). We also consider 
the equation of state of pure neutron matter $\bar E_n(k_n)$ and the asymmetry 
energy $A(k_f)$ in that approach. The downward bending of these quantities
above nuclear matter saturation density seems to be a generic feature of 
perturbative chiral pion-nucleon dynamics.      
\end{abstract}

\bigskip


\bigskip

\bigskip
\section{Introduction}
The present status of the nuclear matter problem is that a quantitatively
successful description can be achieved, using advanced many-body techniques
\cite{akmal}, in a non-relativistic framework when invoking an adjustable
three-body force. Alternative relativistic mean-field approaches, including
non-linear terms with adjustable parameters or explicitly density-dependent
point couplings, are also widely used for the calculation of nuclear matter
properties and finite nuclei \cite{ring}.

In recent years a novel approach to the nuclear matter problem based on
effective field theory (in particular chiral perturbation theory) has emerged
\cite{lutz,nucmat,pot}. The key element there is a separation of long- and 
short-distance dynamics and an ordering scheme in powers of small momenta. At 
nuclear matter saturation density the Fermi-momentum $k_{f0}$ and the pion mass
$m_\pi$ are comparable  scales ($k_{f0}\simeq 2m_\pi$), and therefore pions
must be included as explicit degrees of freedom in the description of the 
nuclear many-body dynamics. The contributions to the energy per particle of
isospin-symmetric nuclear matter $\bar E(k_f)$ originating from chiral
pion-nucleon dynamics have been calculated up to three-loop order in
refs.\cite{lutz,nucmat}. Both calculations include the $1\pi$-exchange
Fock-diagram and the iterated $1\pi$-exchange Hartree- and Fock-diagrams. In
ref.\cite{nucmat} irreducible $2\pi$-exchange is also taken into account and a
momentum cut-off $\Lambda$ is used to regularize the few divergent parts
associated with chiral $2\pi$-exchange. The resulting cut-off dependent
contribution to $\bar E(k_f)$ is completely equivalent to that of a zero-range
NN-contact interaction (see eq.(15) in ref.\cite{nucmat}). At that point the
work of Lutz et al.\,\cite{lutz} deviates and it follows a different
strategy. Two zero-range NN-contact interactions (acting in $^3S_1$ and $^1S_0$
NN-states) proportional to the parameters $g_0+g_A^2/4$ and $g_1+g_A^2/4$ are
introduced (see eq.(4) in ref.\cite{lutz}). The components proportional to 
$g_A^2/4$ cancel the zero-range contribution generated by the $1\pi$-exchange 
Fock-diagram. The other components proportional to $g_0$ and $g_1$ are 
understood to subsume all non-perturbative short-range NN-dynamics relevant 
at densities around nuclear matter saturation density $\rho_0$. In order to be 
consistent with this interpretation the NN-contact vertices proportional to 
$g_{0,1}$ are allowed to occur only in first order. Furthermore, according to 
ref.\cite{nnlutz} pions can be treated perturbatively (at least) in the 
$^1S_0$ partial-wave of NN-scattering if the zero-range pieces they generate 
are removed order by order. Therefore, the NN-contact vertex proportional to 
$g_A^2/4$ occurs also in higher orders (see Fig.1 in ref.\cite{lutz} which
includes diagrams with "filled circle" and "open circle" vertices).    

Despite their differences in the treatment of the effective short-range 
NN-dynamics both approaches \cite{lutz,nucmat} are able to reproduce correctly 
the empirical nuclear matter properties (saturation density $\rho_0$, binding 
energy per particle $-\bar E(k_{f0})$ and compressibility $K$) by adjusting 
only one parameter, either the coupling $g_0+g_1 \simeq 3.23$ or the cut-off
$\Lambda \simeq 0.65\,$GeV. Note that in dimensional regularization all
diagrams evaluated in ref.\cite{lutz} are finite. In the chiral approach of the
Munich group \cite{nucmat,pot} the asymmetry energy $A(k_f)$, the energy per
particle of pure neutron matter $\bar E_n(k_n)$ as well as the (complex)
single-particle potential $U(p,k_f)+i\,W(p,k_f)$ below the Fermi-surface
($p\leq k_f$) have been calculated. Good results (in particular for the
asymmetry energy, $A(k_{f0})=33.8$\,MeV, and the depth of the single-particle
potential, $U(0,k_{f0}) = -53.2\,$MeV) have been obtained with the single
cut-off scale $\Lambda \simeq 0.65\,$GeV adjusted to the binding energy per
particle $-\bar E(k_{f0}) = 15.3\,$MeV. Moreover, when extended to finite
temperatures \cite{liquidgas} this approach reproduces the liquid-gas phase
transition of isospin-symmetric nuclear, however, with a too high value of the 
critical temperature $T_c = 25.5\,$MeV.  

It is the purpose of this work to investigate in the approach of Lutz et 
al.\,\cite{lutz} the single-particle potential $U(p,k_f)+i\, W(p,k_f)$ in
isospin-symmetric nuclear matter as well as the neutron matter equation of
state $\bar E_n(k_n)$ and the asymmetry energy $A(k_f)$. One of our major
conclusions will be that any strong short-range NN-dynamics of non-pionic 
origin should be kept at the mean-field level. It should not be further 
iterated with pion-exchange in contrast to the prescription of power-counting 
rules.  

\section{Nuclear matter equation of state}
\begin{figure}
\begin{center}
\includegraphics[scale=1.5]{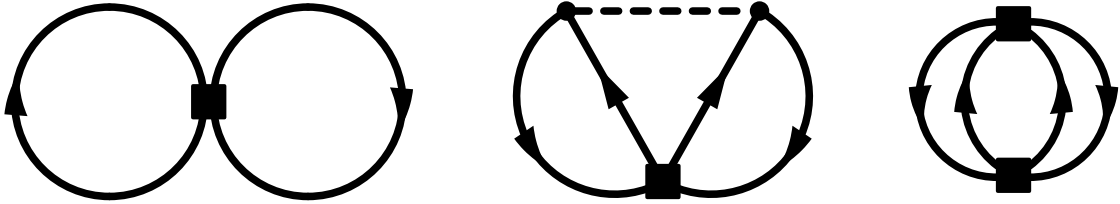}
\end{center}\vspace{-0.2cm}
\caption{Additional in-medium diagrams generated by the NN-contact interactions
introduced in ref.\cite{lutz}. The two NN-contact interactions proportional to
$\gamma+1$ and $\gamma_n+1$ are symbolized by the filled square vertex. The
last diagram is to be understood such that quadratic terms (such as $\gamma^2,
\gamma\gamma_n$ and $\gamma_n^2$) are omitted.}
\end{figure}

Let us first reconsider the equation of state of isospin-symmetric nuclear 
matter as it follows from the calculation of ref.\cite{lutz}. Even though all
contributions to the energy per particle $\bar E(k_f)$ have been given
explicitly in ref.\cite{lutz} we prefer to write down again the extra terms 
generated by the NN-contact interactions proportional to $g_{0,1}+g_A^2/4$ 
(using a more compact notation). The first diagram in Fig.\,1 gives rise to a
contribution to energy per particle of the form:
\begin{equation} \bar E(k_f) = - {(\gamma+1) g_A^2 k_f^3 \over (4\pi f_\pi)^2}
\,, \end{equation}
where we have introduced (for notational convenience) the coefficient $\gamma$
by the relation $(\gamma+1)g_A^2/2= g_0 +g_1+g_A^2/2$. In the second and third 
diagram in Fig.\,1 the contact-interaction proportional to $\gamma+1$ is 
iterated with $1\pi$-exchange or with itself (dropping the $\gamma^2
$-contribution). Putting a medium-insertion\footnote{This is a technical 
notation for the difference between the in-medium and vacuum nucleon 
propagator. For further details, see section 2 in ref.\cite{nucmat}.} at each
of two nucleon propagators with equal orientation one gets:
\begin{equation} \bar E(k_f) = {3(\gamma+1) g_A^4 M m_\pi^4 \over 5(8\pi)^3 
f_\pi^4} \bigg[ 11u -{1\over 2u} -(10+8u^2) \arctan 2u +\bigg( {1\over 8u^3}
+{5\over 2u} \bigg) \ln(1+4u^2) \bigg] \,,\end{equation}
with the abbreviation $u=k_f/m_\pi$. One observes that eq.(2) receives no 
contribution from the third diagram in Fig.\,1 since $\int_0^\infty dl\,1$ is 
set to zero in dimensional regularization. The second and third diagram in
Fig.\,1 with three medium-insertions give rise to the following contribution to
the energy per particle:
\begin{equation} \bar E(k_f) = {9g_A^4 M m_\pi^4 \over (4\pi f_\pi)^4 u^3}
\int_0^u dx\,x^2 \int_{-1}^1 dy \, \Big[ 2uxy +(u^2-x^2y^2)H \Big] \bigg[ 
{\gamma+1 \over 2} \ln(1+s^2) -{s^2 \over 4} \bigg] \,, \end{equation}
with the auxiliary functions $H=\ln(u+x y)-\ln(u-x y)$ and $s=xy +\sqrt{u^2-x^2
+x^2y^2}$. In the chiral limit $m_\pi=0$ only the contribution coming from the 
last term, $-s^2/4$, in the second square bracket survives. The corresponding 
double integral $\int_0^u dx\,x^2 \int_{-1}^1 dy\dots $ has the value $2u^7(
\ln 4-11)/105$. The expansion of the energy per particle up to order ${\cal
O}(k_f^4)$ is completed by adding to the terms eqs.(1,2,3) the contributions 
from the (relativistically improved) kinetic energy, from $1\pi$-exchange and
from iterated $1\pi$-exchange written down in eqs.(5-11) of ref.\cite{nucmat}. 
In case of the $1\pi$-exchange contribution (eq.(6) in ref.\cite{nucmat}) we 
neglect of course the small relativistic $1/M^2$-correction of order 
${\cal O}(k_f^5)$.  

\begin{figure}
\begin{center}
\includegraphics[scale=0.55]{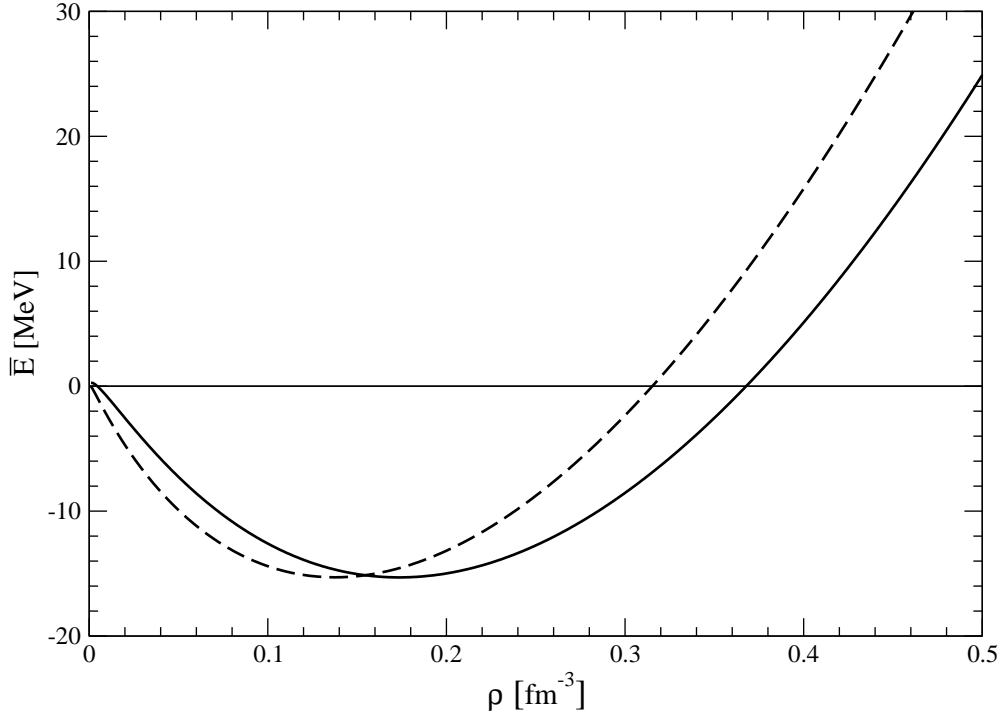}
\end{center}
\caption{The energy per particle of isospin-symmetric nuclear matter $\bar 
E(k_f)$ versus the nucleon density $\rho=2k_f^3/3\pi^2$. The dashed line
corresponds to the approach of ref.\cite{lutz}. The full line results if the
NN-contact interaction is treated in mean-field approximation. In each case 
the coefficient $\gamma$ is adjusted such that the saturation minimum lies at 
$\bar E(k_{f0})=-15.3\,$MeV.} \end{figure}

Now, we have to fix parameters. The pion decay constant $f_\pi=92.4\,$MeV and
the nucleon mass $M=939\,$MeV are well-known. As in ref.\cite{nucmat} we choose
the value $g_A=1.3$. This corresponds via the Goldberger-Treiman relation to a 
$\pi NN$-coupling constant of $g_{\pi N}= g_A M/f_\pi = 13.2$ which agrees with
present empirical determinations of $g_{\pi N}$ from $\pi N$-dispersion 
relation analyses \cite{pavan}. We set $m_\pi = 135\,$MeV (the neutral pion 
mass) since this is closest to the expected value of the pion mass in the
absence of isospin-breaking and electromagnetic effects. 

The dashed line in Fig.\,2 shows the equation of state of isospin-symmetric 
nuclear matter in the approach of ref.\cite{lutz} using the abovementioned 
input parameters. The coefficient $\gamma = 4.086$ has been adjusted such that
the minimum of the saturation curve lies at $\bar E(k_{f0})=-15.3\,$ MeV
\cite{seeger}. The predicted equilibrium density $\rho_0 = 0.138\fmd$
(corresponding to a Fermi-momentum of $k_{f0}=250.1\,$MeV) is somewhat too
low. The same holds for the nuclear matter compressibility $K =k_{f0}^2 \bar 
E''(k_{f0})= 202\,$MeV. Of course, if we use the input parameters of 
ref.\cite{lutz} ($f_\pi = 93\,$MeV, $g_A = 1.26$, $m_\pi= 140\,$MeV and 
$g_0+g_1=3.23$ corresponding to $\gamma =4.07$) we exactly reproduce the 
numerical results of that work. We emphasize that the different treatment of 
the two components of the NN-contact interaction is essential in order to get 
(realistic) nuclear binding and saturation in the framework of ref.\cite{lutz}.
If both components were treated on equal footing in first order (technically 
this is realized by deleting the contribution coming from the third diagram in 
Fig.\,1) the energy per particle $\bar E(k_f)$ would not even develop a 
minimum. 

\section{Real single-particle potential}
Next, we turn to the real part of the single-particle potential $U(p,k_f)$
below the Fermi-surface ($p\leq k_f$) in the framework of ref.\cite{lutz}. As
outlined in ref.\cite{pot} the contributions to $U(p,k_f)$ can be classified 
as two-body and three-body potentials. From the first diagram in Fig.\,1 one
gets a contribution to the two-body potential of the form:   
\begin{equation} U_2(p,k_f) = - {2(\gamma+1) g_A^2 k_f^3 \over (4\pi f_\pi)^2}
\,, \end{equation}
which is just twice its contribution to the energy per particle (see eq.(1)). 
From the second diagram in Fig.\,1 one derives a contribution to the two-body 
potential of the form: 
\begin{eqnarray} U_2(p,k_f) &=& {(\gamma+1) g_A^4 M m_\pi^4 \over (4\pi)^3 
f_\pi^4} \bigg\{u+{1\over 4x}(x^3-3x-3u^2x -2u^3) \arctan(u+x)  \nonumber \\ 
& & +{1\over 4x}(x^3-3x-3u^2x +2u^3) \arctan(u-x)\nonumber \\ && +{1\over 8x}
(1+3u^2-3x^2) \ln{1+(u+x)^2 \over 1+(u-x)^2} \bigg\}\,, \end{eqnarray}
with the abbreviation $x=p/m_\pi$. The second and third diagram in Fig.\,1
give each rise to three different contributions to the three-body potential. 
Altogether, they read:  
\begin{eqnarray} U_3(p,k_f) &=& {3g_A^4 M m_\pi^4 \over (4\pi f_\pi)^4}
\int_{-1}^1 dy \, \bigg\{ \Big[ 2uxy +(u^2-x^2y^2)H\Big] \bigg[ {\gamma+1 \over
2} \ln(1+s^2) -{s^2 \over 4} \bigg]\nonumber \\ 
& &  +\int_{-xy}^{s-xy} d\xi\,\bigg[ 2u \xi +(u^2-\xi^2) \ln{u+\xi \over u-\xi}
\bigg] \,{ (2\gamma+1)(xy+\xi)-(xy+\xi)^3 \over 2[1+(xy+\xi)^2] } \nonumber \\ 
& & +\int_0^u d\xi\, {\xi^2 \over x} \bigg[ (\gamma+1 )\ln(1+\sigma^2) 
-{\sigma^2 \over 2} \bigg]\ln{|x+\xi y| \over |x-\xi y|} \bigg\}\,, 
\end{eqnarray}
with the auxiliary function $\sigma=\xi y +\sqrt{u^2-\xi^2+\xi^2y^2}$. The real
single-particle potential $U(p,k_f)$ is completed by adding to the terms
eqs.(4,5,6) the contributions from $1\pi$-exchange and iterated $1\pi$-exchange
written down in eqs.(8-13) of ref.\cite{pot}. Again, the (higher order) 
relativistic $1/M^2$-correction to $1\pi$-exchange (see eq.(8) in 
ref.\cite{pot}) is neglected for reasons of consistency.   

\begin{figure}
\begin{center}
\includegraphics[scale=0.55]{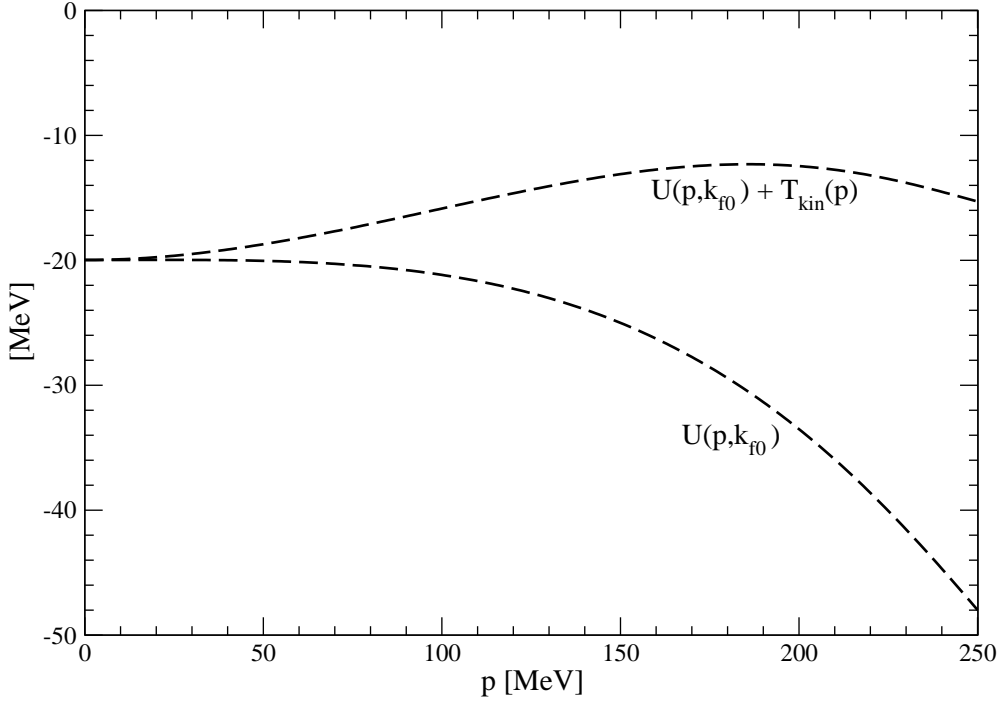}
\end{center}
\caption{The lower curve shows the real part of the single-particle potential 
$U(p,k_{f0})$ at saturation density $k_{f0}=250.1\,$MeV in the approach of Lutz
et al.\,\cite{lutz}. The upper curve includes in addition the relativistically
improved kinetic energy $T_{\rm kin}(p)=p^2/2M-p^4/8M^3$.} 
\end{figure}

The lower curve in Fig.\,3 shows the momentum dependence of the real 
single-particle potential $U(p,k_{f0})$ at saturation density
$k_{f0}=250.1\,$MeV as it arises in the framework of Lutz et al.\,\cite{lutz}. 
The predicted potential depth for a nucleon at the bottom of the Fermi-sea is
only $U(0,k_{f0})=-20.0\,$ MeV. In magnitude this is much smaller than the 
typical depth $U_0 \simeq -53\,$MeV of the empirical optical model potential 
\cite{hodgson} or the nuclear shell model potential \cite{bohr}. The upper 
curve in Fig.\,3 includes the (relativistically improved) single-nucleon
kinetic energy $T_{\rm kin}(p)=p^2/2M-p^4/8M^3$. As required by the
Hugenholtz-van-Hove theorem \cite{vanhove} this curve ends at the Fermi-surface
$p=k_{f0}$ with the value $\bar E(k_{f0})=-15.3\,$MeV. A further important
check is provided by the sum rule for the two- and three-body potentials
$U_{2,3}(p,k_f)$ written down in eq.(5) of ref.\cite{pot}. It holds with very
high numerical accuracy in the present calculation. 

The momentum dependence of the two (dashed) curves in Fig.\,3 is completely
unrealistic. Most seriously, the total single-particle energy $T_{\rm kin}(p)+
U(p,k_{f0})$ (upper curve) does not rise monotonically with the nucleon
momentum $p$, but instead it starts to bend downward above $p\simeq
190\,$MeV. This implies a negative effective nucleon mass at the Fermi-surface,
$M^*(k_{f0}) \simeq-3.5\,M$, and a negative density of states with dramatic
consequences for the finite temperature behavior of nuclear matter. Because of
such pathological features of the underlying single-particle potential the
scheme of Lutz et al.\,\cite{lutz} has to be rejected in its present form.  

\begin{figure}
\begin{center}
\includegraphics[scale=0.55]{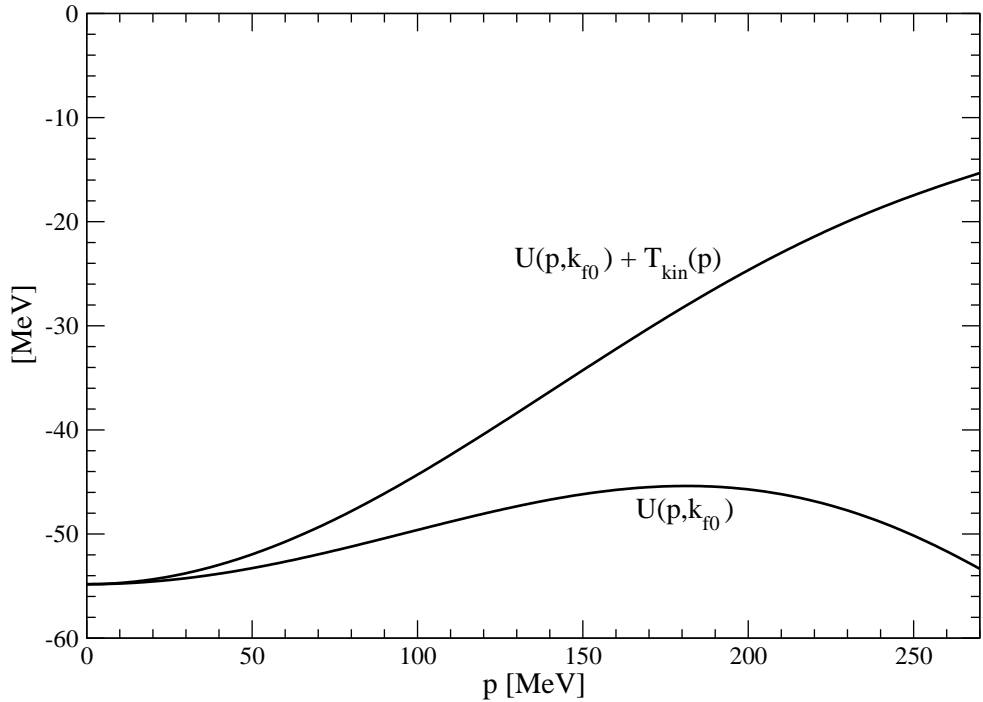}
\end{center}
\caption{The lower curve shows the real part of the single-particle potential 
$U(p,k_{f0})$ at saturation density $k_{f0}=270.3\,$MeV in a mean-field 
treatment of the NN-contact interaction. }
\end{figure}

The overly strong momentum dependence of $U(p,k_{f0})$ comes from the second 
and third diagram in Fig.\,1 in which the NN-contact interaction proportional 
to the large coefficient $\gamma+1$ is further iterated.  We propose to drop
these three-loop diagrams and to keep the NN-contact interaction (of 
unspecified dynamical origin) at the mean-field level. The resulting equation
of state obtained by leaving out the contributions eqs.(2,3) and adjusting
$\gamma = 6.198$ is shown by the full line in Fig.\,2. The predicted saturation
density is now $\rho_0 = 0.174\fmd$ (corresponding to a Fermi-momentum of
$k_{f0}=270.3\,$MeV) and the nuclear matter compressibility has the value $K =
253\,$MeV. Note that the scheme of ref.\cite{lutz} modified by a 
mean-field treatment of the NN-contact interaction becomes equivalent to the
truncation at fourth  order in small momenta of our previous work
\cite{nucmat,pot} after the identification of parameters, $\gamma+1 = 10 g_A^2
\Lambda M/(4\pi f_\pi)^2$, with $\Lambda$ denoting the cut-off scale.   

The lower full curve in Fig.\,4 shows the momentum dependence of the
real single-particle potential at saturation density $k_{f0}=270.3\,$MeV
which results in a mean-field approximation of the NN-contact interaction (by
leaving out the contributions eqs.(5,6)). The predicted potential depth
$U(0,k_{f0}) = -54.8\,$MeV is in good agreement with that of optical model
\cite{hodgson} or nuclear shell model potentials \cite{bohr}. Most importantly,
the total single-particle energy $T_{\rm kin}(p)+ U(p,k_{f0})$ (upper curve)
grows now monotonically with the nucleon momentum $p$, as it should. The up-
and downward bending of the lower full curve in Fig.\,4 is however still too
strong. The negative slope of $U(p,k_{f0})$ at the Fermi-surface $p=k_{f0}$
leads to a too large effective nucleon mass $M^*(k_{f0}) \simeq 2.9\,M$ which
reflects itself in a too high critical temperature $T_c \simeq 25\,$MeV of the
liquid-gas phase transition \cite{liquidgas}. More elaborate calculation of
nuclear matter in effective (chiral) field theory are necessary in order to
cure this problem of the too large effective nucleon mass $M^*(k_{f0})$.  

\section{Imaginary single-particle potential}
In this section, we discuss the imaginary part of the single-particle potential
$W(p,k_f)$ for $p \leq k_f$ as it arises in the scheme of Lutz et 
al.\,\cite{lutz}. This quantity determines the half-width of nucleon-hole
states in the Fermi-sea. As outlined in ref.\cite{pot} the contributions to
$W(p,k_f)$ can be classified as two-body, three-body and four-body terms. From
the second and third diagrams in Fig.\,1 one derives a two-body term of the
form:    
\begin{eqnarray} W_2(p,k_f) &=& {g_A^4 M m_\pi^4 \over (8\pi)^3 f_\pi^4} 
\bigg\{u^2x^2 +{3u^4\over 2} -{x^4 \over 10} +(\gamma+1)\bigg[ 4+14u^2 -{22x^2 
\over 3}\nonumber \\ & &  +{2\over x} (3x^2-3u^2-1) \Big[\arctan(u+x)-
\arctan(u-x) \Big] \nonumber \\ & &  +{1\over x} (x^3-3x-3u^2x -2u^3)
\ln[1+(u+x)^2]\nonumber \\ & &  +{1\over x} (x^3-3x-3u^2x +2u^3) \ln[1+(u-x)^2]
\bigg]\bigg\}\,. \end{eqnarray} 
The associated three-body term reads: 
\begin{eqnarray} W_3(p,k_f) &=& {3\pi g_A^4 M m_\pi^4 \over (4\pi f_\pi)^4}
\int_{-1}^1 dy \, \bigg\{(\gamma+1) \bigg[ 2xy(s-\arctan s) -{s^2 \over 2} 
\nonumber \\ & & +\Big( {1\over 2}+u^2-x^2y^2\Big)\ln(1+s^2)\bigg] + s^2 \bigg(
{s^2 \over 8} -{u^2 \over 2} -{sxy \over 3} +{x^2 y^2 \over 2} \bigg) \nonumber
\\ & & +\int_0^u d\xi\, {\xi^2 \over x} \theta(x-\xi|y|) \bigg[ (\gamma +1)\ln
(1+\sigma^2) -{\sigma^2 \over 2} \bigg]\bigg\}\,, \end{eqnarray}
and the four-body term is given by the expression: 
\begin{eqnarray} W_4(p,k_f) &=& {3\pi g_A^4 M m_\pi^4 \over (4\pi f_\pi)^4}
\bigg\{ (\gamma+1) \bigg[ {4x^2 \over 3} -1 +\ln(1+4x^2) +\bigg( {1\over
2x}-2x\bigg) \arctan2x \bigg]\nonumber \\ & & -{4x^4 \over 15} +\int_{-1}^1 
dy  \, \bigg[s^2 \bigg( {u^2\over 2}-{s^2\over 4}+{2sxy \over 3}
-{x^2y^2 \over 2} \bigg)\nonumber \\ & &+(\gamma+1) \Big[ 4xy(\arctan s-s) 
+s^2 +(x^2y^2-1-u^2)\ln(1+s^2) \Big] \bigg]  \bigg\}\,. \end{eqnarray}
The additional contributions from the iterated $1\pi$-exchange Hartree- and
Fock-diagram are collected in eqs.(20-25) of ref.\cite{pot}. 
The total imaginary single-particle potential evaluated at zero nucleon
momentum ($p=0$) can even be written as a closed form expression: 
\begin{eqnarray}  W(0,k_f) &=&{3\pi g_A^4M m_\pi^4 \over (4\pi f_\pi)^4}\bigg\{
{u^4\over 2}+(\gamma-2) u^2-{2u^2\over 1+u^2}+{\pi^2\over 12}
+{\rm Li}_2(-1-u^2) \nonumber\\ && + \Big[4-\gamma +\ln(2+u^2)-{1\over 2}
\ln(1+u^2) \Big] \ln(1+u^2) \bigg\} \,, \end{eqnarray}
where Li$_2(-a^{-1}) = \int_0^1 d\zeta \, (\zeta+a)^{-1} \ln\zeta$ denotes the 
conventional dilogarithmic function.

The dashed line in Fig.\,5 shows the momentum dependence of the imaginary 
single-particle potential $W(p,k_{f0})$ at saturation density
$k_{f0}=250.1\,$MeV as it arises in the approach of ref.\cite{lutz}. The
predicted value $W(0,k_{f0})=51.1$\,MeV lies outside the range $20-40\,$MeV
obtained in calculations based on (semi)-realistic NN-forces
\cite{grange,schuck}. The full line in Fig.\,5 corresponds to a mean-field
approximation of the NN-contact interaction. Up to a slight change in the
equilibrium Fermi-momentum $k_{f0}=270.3\,$MeV the full curve in Fig.\,5 agrees
with the one shown in Fig.\,4 of ref.\cite{pot}. The considerably reduced 
value $W(0,k_{f0})=28.4$\,MeV indicates the large contribution of the iterated
diagrams in Fig.\,1 to the imaginary single-particle potential $W(p,k_f)$. 
Note that both curves in Fig.\,5 vanish quadratically near the Fermi-surface as
required by Luttinger's theorem \cite{luttinger}. As further check on our
calculation we verified the zero sum rule: $\int_0^{k_f} dp\,p^2 [6 W_2(p,k_f)
+4 W_3(p,k_f)+3W_4(p,k_f)]=0$, for the two-, three- and four-body components
$W_{2,3,4}(p,k_f)$ written in eqs.(7,8,9).  

\begin{figure}
\begin{center}
\includegraphics[scale=0.55]{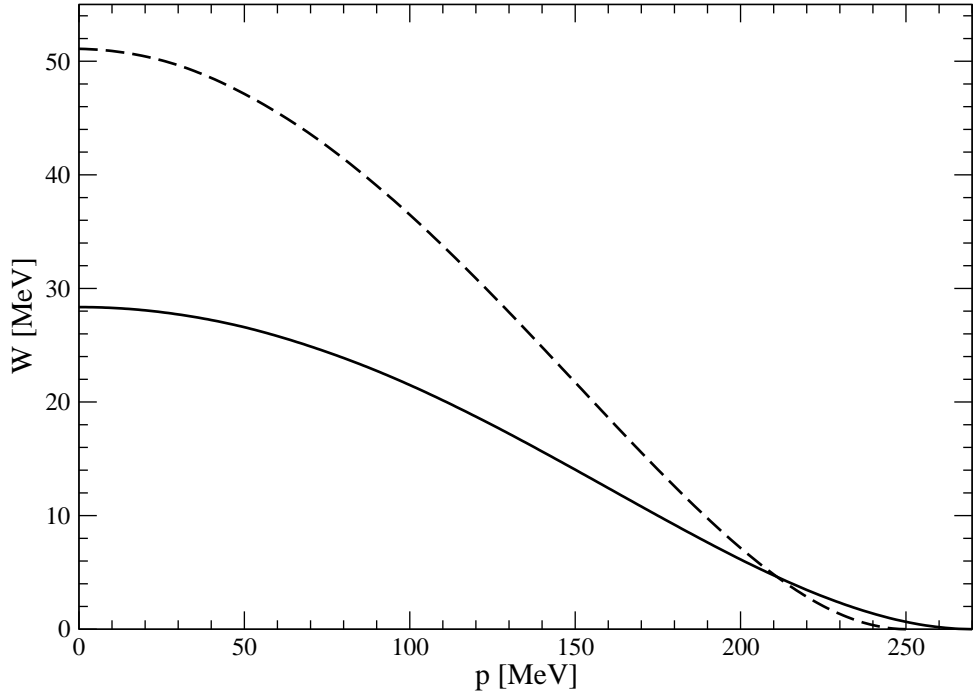}
\end{center}
\caption{The imaginary part of the single-particle potential $W(p,k_{f0})$ at
saturation density versus the nucleon momentum $p$. The dashed line corresponds
to the approach of Lutz et al.\,\cite{lutz} and the full line shows the result
obtained in a mean-field treatment of the NN-contact interaction.}
\end{figure}

\section{Neutron matter}
In this section we discuss the equation of state of pure neutron matter. In 
the scheme of Lutz et al.\,\cite{lutz} the energy per particle of pure neutron 
matter $\bar E_n(k_n)$ depends exclusively on the coefficient $g_1$ 
parameterizing the short-range NN-interaction in the channel with total isospin
$I=1$. There is no need to write down explicitly the contributions of the 
diagrams in Fig.\,1 to $\bar E_n(k_n)$. These expressions are easily obtained
from eqs.(1,2,3) by replacing $k_f$ by the neutron Fermi-momentum $k_n$,
by replacing the coefficient $\gamma$ by a new one $\gamma_n$, and by
multiplying the formulas with a relative isospin factor $1/3$. The relation
$(\gamma_n+1) g_A^2/4= g_1+ g_A^2/4$ defines this new coefficient $\gamma_n$. 
The additional contributions to $\bar E_n(k_n)$ from the kinetic energy,
$1\pi$-exchange and iterated $1\pi$-exchange are written down in eqs.(32-37) of
ref.\cite{nucmat} (neglecting again the relativistic $1/M^2$-correction to
$1\pi$-exchange).      

\begin{figure}
\begin{center}
\includegraphics[scale=0.55]{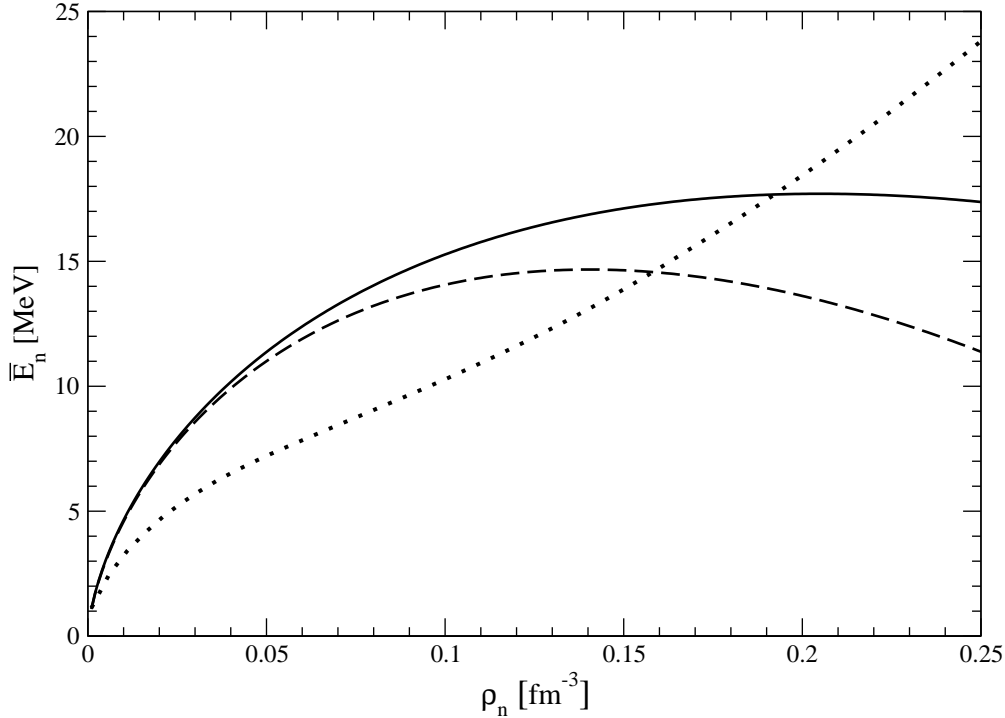}
\end{center}
\caption{The energy per particle of pure neutron matter $\bar E_n(k_n)$ versus
the neutron density $\rho_n = k_n^3/3\pi^2$. The dashed line corresponds
to the approach of Lutz et al.\,\cite{lutz} and the full line shows the result
obtained in mean-field approximation of the $nn$-contact interaction. The
dotted line stems from the many-body calculation of the Urbana group
\cite{urbana}.} 
\end{figure}

The dashed line in Fig.\,6 shows the energy per particle of pure neutron matter
$\bar E_n(k_n)$ versus the neutron density $\rho_n = k_n^3/3\pi^2$ as it arises
in the approach of ref.\cite{lutz}. The coefficient $\gamma_n = 0.055$ has been
adjusted to the empirical value of the asymmetry energy $A(k_{f0}=250.1\,{\rm
MeV}) = 33.2\,$MeV (see next section). The downward bending of the dashed 
curve in Fig.\,6 above $\rho_n >0.15\fmd$ is even stronger than in our previous
work \cite{nucmat} (see Fig.\,8 therein). This property can be understood by 
taking the chiral limit ($m_\pi \to 0$) of the calculated neutron matter 
equation of state and considering the coefficient $\beta_n$ in front of the 
term $k_n^4/M^3$. In the approach of Lutz et al.\,\cite{lutz} one has: 
\begin{equation} \beta_n =  - {1\over 70} \Big({g_{\pi N}\over 4\pi} \Big)^4
(4\pi^2 +17+16\ln2) -{3\over 56}=-1.23\,, \end{equation}
 which is 2.2 times 
the negative value of $\beta_n$ found in ref.\cite{nucmat}. The full line in
Fig.\,6 shows the equation of state of pure neutron matter obtained in 
mean-field approximation of the $nn$-contact interaction proportional to
$\gamma_n+1$ after adjusting $\gamma_n = 0.788$ to the empirical value of the
asymmetry energy $A(k_{f0}=270.3\,{\rm MeV}) = 33.2\,$MeV. The downward bending
of the full curve in Fig.\,6 is weaker and it sets in at somewhat higher
densities $\rho_n >0.2\fmd$. The dotted line in Fig.\,6 stems from the
many-body calculation of the Urbana group \cite{urbana}. This curve should be
considered as a representative of the host of existing realistic neutron matter
calculations which scatter around it. The systematic deviations observed in 
Fig.\,6 indicate that the neutron matter equation of state of ref.\cite{nucmat}
cannot be improved by a different treatment of the short-range NN-dynamics
alone. The downward bending of $\bar E_n(k_n)$ above $\rho_n >0.2\fmd$ seems to
be a generic feature of perturbative chiral $\pi N$-dynamics truncated at
three-loop order. 
\section{Asymmetry energy}
Finally, we turn to the density dependent asymmetry energy $A(k_f)$ in the
approach of ref.\cite{lutz}. The asymmetry energy is generally defined by the 
expansion of the energy per particle of isospin-asymmetric nuclear matter
(described by different proton and neutron Fermi momenta $k_{p,n}= k_f(1\mp
\delta)^{1/3}$) around the symmetry line: $\bar E_{\rm as}(k_p,k_n) = \bar
E(k_f)+\delta^2\,A(k_f) +{\cal O}(\delta^4)$. Evaluation of the first diagram
in Fig.\,1 leads to the following contribution to the asymmetry energy:   
\begin{equation} A(k_f) = {g_A^2 k_f^3 \over 3(4\pi f_\pi)^2}(3\gamma-
2\gamma_n+1)   \,, \end{equation}
with the coefficients $\gamma = 2(g_0+g_1)/g_A^2$ and $\gamma_n = 4g_1/g_A^2$ 
in the notation of ref.\cite{lutz}. Putting a medium-insertion at each of two
nucleon propagators with equal orientation one gets from the second and third
diagram in Fig.\,1: 
\begin{eqnarray} A(k_f) &=& {g_A^4 M m_\pi^4 \over 3(8\pi)^3 f_\pi^4} \bigg\{ 
2(\gamma+1)u +8(2\gamma-\gamma_n+1)u^2 \arctan 2u \nonumber \\ && 
+\bigg[(2\gamma_n-6\gamma-4) u - {\gamma+1 \over 2u}\bigg] \ln(1+4u^2) \bigg\}
\,.\end{eqnarray} 
The same diagrams with three medium-insertions give rise to the following
contribution to the asymmetry energy:
\begin{eqnarray} A(k_f) &=& {g_A^4 M m_\pi^4 \over (4\pi f_\pi)^4 u^3}
\int_0^u dx\,x^2 \int_{-1}^1 dy\,\Bigg\{\Big[ 2uxy +(u^2-x^2y^2)H \Big] \bigg( 
4ss' -{2\over 3}s'\,^2 -{2\over 3}ss''- {7\over 2}s^2 \bigg) \nonumber \\ && +
(\gamma+1) \bigg\{ \bigg[ {uxy(11u^2-15x^2y^2)\over 3(u^2-x^2y^2)} +{1\over 2}
(u^2-5x^2y^2)H \bigg] \ln(1+s^2)-{4u^2s^2H \over 3(1+s^2)} \nonumber \\ && + 
{2uxy +(u^2-x^2y^2)H \over 6(1+s^2)^2} \Big[ 8s(1+s^2) (3s+s''-5s') +(1-s^2) 
(3s^2-8ss'+8s'\,^2) \Big]\bigg\} \nonumber \\ && + 2u^2(\gamma_n+1)\bigg[{2uxy
\ln(1+s^2) \over 3(u^2-x^2y^2)} +\bigg(  \ln(1+s^2)+{2s^2 \over 3(1+s^2)}
\bigg)H \bigg] \Bigg\}  \,, \end{eqnarray}
with $s' = u\, \partial s/\partial u$ and $s''=u^2\,\partial^2 s/\partial u^2$
denoting partial derivatives. In the chiral limit $m_\pi=0$ only the terms in
the first line of eq.(14) survive. The corresponding double integral $\int_0^u
dx\,x^2 \int_{-1}^1 dy \dots$ has the value $4u^7(\ln 4 -1)/15$. The asymmetry
energy is completed by adding to the terms eqs.(12,13,14) the contributions 
from the kinetic energy, (static) $1\pi$-exchange and iterated $1\pi$-exchange
written down in eqs.(20-26) of ref.\cite{nucmat}.  

The dashed line in Fig.\,7 shows the density dependence of the asymmetry energy
$A(k_f)$ in the approach of Lutz et al.\,\cite{lutz} with the coefficient 
$\gamma_n = 0.055$ adjusted (at fixed $\gamma=4.086$) to the empirical value 
$A(k_{f0}=250.1\,{\rm MeV}) = 33.2\,$MeV \cite{seeger}. The full line in 
Fig.\,7 corresponds to the result obtained in mean-field approximation of the
NN-contact interaction by dropping the contributions eqs.(13,14). In that case
the empirical value $A(k_{f0}=270.3\,{\rm MeV}) = 33.2\,$MeV \cite{seeger} is 
reproduced by tuning (at fixed $\gamma = 6.198$) the coefficient $\gamma_n$ to 
the value $\gamma_n = 0.788$. Both curves in Fig.\,7 behave rather similarly. 
In each case the asymmetry $A(k_f)$ reaches it maximum close to the respective
saturation density $\rho_0$ and then it starts to bend downward. Since the 
same (unusual) feature has also been observed in ref.\cite{nucmat} it seems to
be generic for perturbative chiral $\pi N$-dynamics truncated at three-loop
order. 

\begin{figure}
\begin{center}
\includegraphics[scale=0.55]{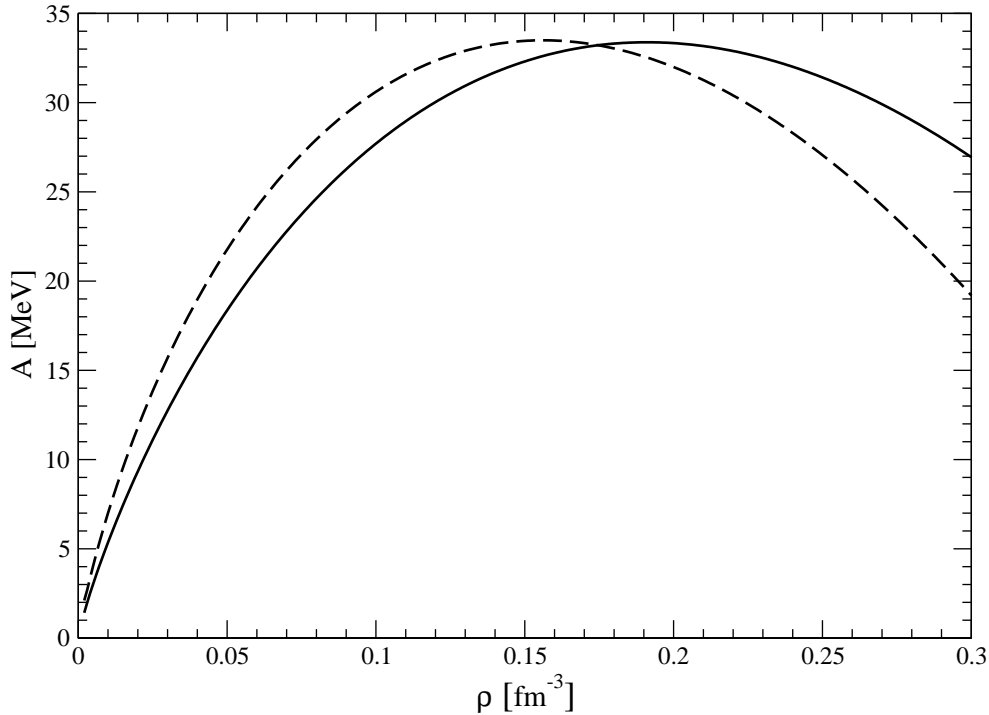}
\end{center}
\caption{The asymmetry energy $A(k_f)$ versus the nucleon density $\rho=2k_f^3
/3\pi^2$. The dashed line corresponds to the approach of Lutz et
al.\,\cite{lutz} and the full line shows the result obtained in mean-field
approximation of the NN-contact interactions. The parameter $\gamma_n$ is in
each case adjusted to the (empirical) value  $A(k_{f0})=33.2\,$MeV
\cite{seeger}.} 
\end{figure}

\section{Concluding remarks}
In this work have we continued and extended the chiral approach to nuclear 
matter of Lutz et al.\,\cite{lutz} by calculating the underlying 
single-particle potential. The potential for a nucleon at the bottom of the
Fermi-sea $U(0,k_{f0})=-20.0\,$MeV is not deep enough. Most seriously, the
total single-particle energy $T_{\rm kin}(p)+ U(p,k_{f0})$ does not grow
monotonically with the nucleon momentum $p$. The thereof implied negative
effective nucleon mass at the Fermi-surface $M^*(k_{f0}) \simeq -3.5\,M$ and
the negative density of states will ruin the behavior of nuclear matter at
finite temperatures. The half-width of nucleon-holes at the bottom of the
Fermi-sea  $W(0,k_{f0})=51.1\,$MeV comes also out too large in that approach. A
good nuclear matter equation of state and better (but still not yet optimal)
single-particle properties can be obtained if the NN-contact interaction
(proportional to the coefficient $g_0+g_1+g_A^2/2$) is kept at the mean-field
level and not further iterated. The energy per particle of pure neutron matter
$\bar E_n(k_n)$ and the asymmetry energy $A(k_f)$ depend on a second parameter
$g_1$ in the scheme of ref.\cite{lutz}. Their density dependence is similar to
the results of the one-parameter calculation in ref.\cite{nucmat}. The downward
bending of $\bar E_n(k_n)$ and $A(k_f)$  above saturation density $\rho_0$
(less pronounced if the NN-contact interaction is kept at mean-field level)
seems to be generic for perturbative chiral $\pi N$-dynamics. More elaborate
calculations of nuclear matter in effective (chiral) field theory which fulfill
all (semi)-empirical constraints are necessary.          
\section*{Acknowledgment}
We thank M. Lutz for useful discussions. 

\end{document}